\documentclass[preprint,aps,prl,endfloats]{revtex4-1}

\usepackage{graphicx}
\usepackage{dcolumn}
\usepackage{bm}
\usepackage{amsmath}
\usepackage{latexsym}
\usepackage{hyperref}
  
\newcommand{\rfig}[1]{Fig.~\ref{#1}}
\newcommand{\rtab}[1]{Table~\ref{#1}}
\newcommand{\rref}[1]{Ref.~\onlinecite{#1}}
\newcommand{\be}{\begin{equation}}
\newcommand{\ee}{\end{equation}}
\newcommand{\req}[1]{Eq.~(\ref{#1})}

\begin{document}

\title{Observation of vacancy-induced suppression of electronic cooling in defected graphene}

\author{Qi Han}
\thanks{These two authors contributed equally.}
\author{Yi Chen}
\thanks{These two authors contributed equally.}
\author{Gerui Liu, Dapeng Yu}
\author{Xiaosong Wu}
\email{xswu@pku.edu.cn}
\affiliation{
State Key Laboratory for Artificial Microstructure and Mesoscopic Physics, Peking University, Beijing 100871, China \\
Collaborative Innovation Center of Quantum Matter, Beijing 100871, China}


\begin{abstract}
Previous studies of electron-phonon interaction in impure graphene have found that static disorder can give rise to an enhancement of electronic cooling. We investigate the effect of dynamic disorder and observe over an order of magnitude suppression of electronic cooling compared with clean graphene. The effect is stronger in graphene with more vacancies, confirming its vacancy-induced nature. The dependence of the coupling constant on the phonon temperature implies its link to the dynamics of disorder. Our study highlights the effect of disorder on electron-phonon interaction in graphene. In addition, the suppression of electronic cooling holds great promise for improving the performance of graphene-based bolometer and photo-detector devices. 
\end{abstract}


\maketitle
\section{Introduction}

In recent years, there has been considerable interest in utilizing graphene as photo-detectors\cite{Xu2009,Gabor2011,Kalugin2011,Sun2012,Yan2012,Vora2012,Yan2012a,Cai2013}. Most of these detectors are based on a hot electron effect, \textit{i.e.} the electronic temperature being substantially higher than the lattice temperature. Two properties of graphene strongly enhance the effect. First, low carrier density gives rise to a very small electron specific heat. Second, weak electron-phonon (e-p) interaction reduces the heat transfer from the electron gas to the lattice. Thus, it is of practical interest to understand the e-p interaction in graphene. Both theoretical and experimental efforts have been devoted to this topic. Earlier work was mainly focused on clean graphene and considered the Dirac spectrum of electrons\cite{Kubakaddi2009,Bistritzer2009,Tse2009,Viljas2010,Betz2012,Baker2012,Baker2013}. As the important role of impurities in electronic transport has been revealed, its effects on the e-p interaction began to draw attention\cite{Vasko2011,Chen2012,Song2012}. For instance, due to the chiral nature of electrons, long range and short range potentials scatter electrons differently in graphene\cite{McCann2006,Nomura2006a,Wu2007}. Recently, a strong enhancement of electronic cooling via e-p interaction in presence of short range disorder has been predicted\cite{Song2012}. This is achieved via a so-called supercollision process. When the carrier density is low, the Bloch-Gr\"uneisen temperature $T_\text{BG}$ can be quite small. Since $T_\text{BG}$ sets the maximum wave vector of phonons that can exchange energy with electrons, when $T_\text{BG} < T $, only a portion of phonons can contribute to the energy relaxation. Interestingly, in presence of short range potentials, the theory has found that a disorder-assisted scattering process can occur, in which all available phonons are able to participate. As a result, the energy relaxation is strongly enhanced. Shortly, two experiments confirmed the supercollision\cite{Betz2013,Graham2013}, although long range potential scattering usually dominates in such samples\cite{Adam2007,Chen2008}. In the case of long range potentials, Chen and Clerk have also predicted an increase of electronic cooling at low temperature for weak screening\cite{Chen2012}. Note that besides the different potential profiles, \textit{e.g.} long range or short range, disorder can be static or dynamic. Despite these studies, in which only static disorder was considered, the dynamics of disorder has not been addressed.

Here, we present an experimental investigation of the effect of vacancy on electronic cooling in both monolayer and bilayer defected graphene. In contrast to typical scattering potentials previously treated in theories or encountered in experiments, which are static, vacancies in our defected graphene are dragged by phonons, hence highly dynamic. By studying the nonlinear electric transport of defected graphene, a strong \textit{suppression} of e-p energy relaxation, instead of an \textit{enhancement} in the case of static potentials, has been observed. The more disordered the graphene film, the stronger the suppression is. Our work provides new experimental insight on the effect of scattering potential on e-p interaction. Moreover, the suppression suggests that the performance of graphene hot electron photo-detectors can be further improved by introducing vacancies.

\section{Experiment}
In this work, we have investigated four exfoliated graphene samples on Si/SiO$_2$ substrates. Thickness of all the monolayer (SM1 and SM2) and bilayer (SB1, SB2) samples were estimated by optical contrast and confirmed by Raman spectroscopy\cite{Ferrari2006}. Graphene flakes were patterned into ribbons, using e-beam lithography. 5 nm Ti/80 nm Au were e-beam deposited, followed by lift-off to form electrodes. Typical sample geometry can be seen in the inset of \rfig{rt}a. In order to introduce vacancies, samples were then loaded into a Femto plasma system and subject to Argon plasma treatment for various periods (from 1 to 5 s)\cite{Chen2013}. Four-probe electrical measurements were carried out in a cryostat using a standard lock-in technique. Room temperature $\pi$-filters were used to avoid heating of electrons by radio frequency noise. Information for four samples are summarized in \rtab{tab:info}. 
\begin {table*}[htb]
\caption {\label {tab:info} Sample information of four investigated devices. Different Ar gas flow rates and plasma treatment times have been applied to produce different amount of vacancies. ${V_\text{CNP}}$ is the charge neutrality point (CNP) of samples and $\xi$ is the localization length near the CNP.}
\begin {ruledtabular}
\begin {tabular}{ccccccc}
Devices & Length($\mu$m) & Width($\mu$m) & Ar flow rate(sccm) & Plasma treatment period(s) & ${V_\text{CNP}}\rm(V)$ & $\xi\rm(nm) $ \\
\hline
SM1 & 2    & 3   & 3  & 1   & 14.5 & 156\\
SM2 & 6.7  & 2.7 & 4  & 3   & 30   & 21\\
SB1 & 3    & 2.7 & 4  & 3.5 & 70   & 50\\
SB2 & 6    & 2.7 & 4  & 5   & 57   & 54\\
\end {tabular}
\end {ruledtabular}
\end {table*}

\section{Results and discussion}

\begin{figure}[htb]
\includegraphics[width=1\textwidth]{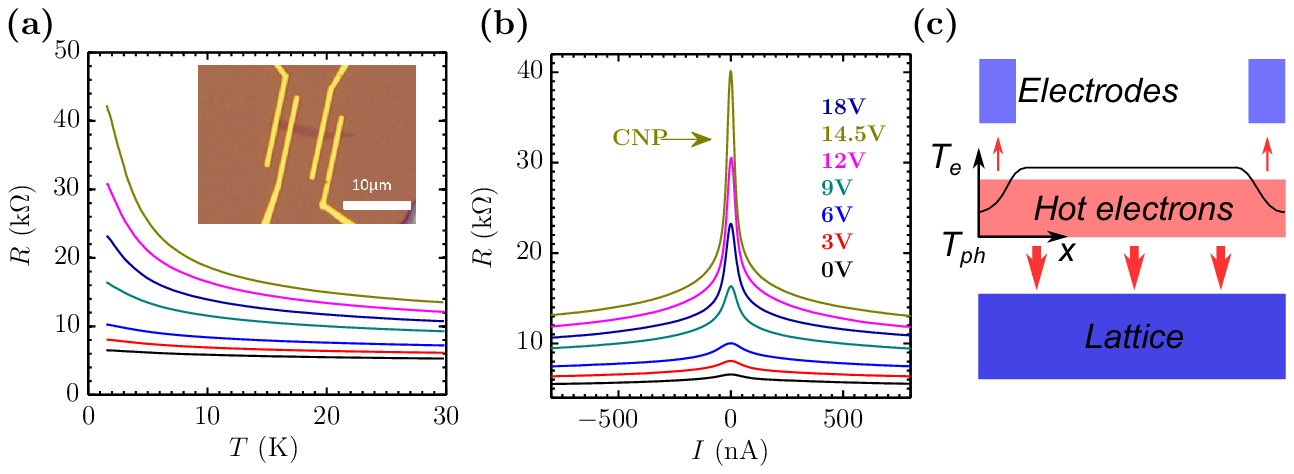}
\caption{\label{rt} Resistance of defected graphene. (a) Temperature dependence of resistance in sample SM1 at different gate voltages, showing divergence at low temperature. Inset: Optical micrograph of a typical device configuration. (b) Resistance of SM1 as a function of Joule heating current at different gate voltages at $T=1.5$ K. The CNP is at 14.5 V. (c)Thermal model for the structure. The pathways of heat dissipation are indicated by red arrows.}
\end{figure}

Previously, we have already demonstrated a hot electron bolometer based on disordered graphene\cite{Han2013}. It has been shown that the divergence of the resistance at low temperature can be utilized as a sensitive thermometer for electrons. By applying Joule heating, the energy transfer rate between the electron gas and the phonon gas can be obtained. The same method has been employed in this work. As showing in \rfig{rt}a, the resistance of defected graphene exhibits a sharp increase as the temperature decreases. The divergence becomes stronger as one approaches the CNP. The $R-T$ behavior can be well fitted to variable range hopping transport, described as $R \propto \exp[(T_0/T)^{1/3}]$\cite{Mott1968}. Here, the characteristic temperature $T_0=12/[\pi k_\text{B} \nu(E_\text{F}) \xi^2]$, with $k_\text{B}$ the Boltzmann constant, $\nu(E_\text{F})$ the density of states at the Fermi level $E_\text{F}$, and $\xi$ the localization length. By fitting to this formula, the localization length $\xi$ is determined. It is employed as a measure of the degree of disorder. $\xi$ near the CNP for all samples are listed in \rtab{tab:info}.

In the steady state of Joule heating, the electron cooling power equals to the heating power. The corresponding thermal model is sketched in \rfig{rt}c. Two thermal energy transfer pathways are indicated, \textit{i.e.} via electron diffusion into electrodes or e-p interaction into the lattice. In our strongly disordered graphene, the former is significantly suppressed due to a very low carrier diffusivity. It has been found that e-p interaction dominates the energy dissipation in such devices\cite{Han2013}. Then, the electronic temperature can be directly inferred from the resistance. Furthermore, it is estimated that the thermal conductance between the graphene lattice and the substrate is much higher than that due to e-p interaction. Thus, the phonon temperature $T_\text{ph}$ is approximately equal to the substrate temperature $T$\cite{Yan2012,Betz2013,Borzenets2013}. Under these conditions, the energy balance at the steady state of Joule heating can be written as
\be
\label{eq1}
P=A(T_\text{e}^\delta-T_\text{ph}^\delta)
\ee
where $P$ is the Joule Heating power, $A$ is the coupling constant and $T_\text{e}$ is the electronic temperature. $\delta$ ranges from 2 to 6, depending on the detail of the e-p scattering process\cite{Viljas2010}.

\begin{figure}[htb]
\includegraphics[width=1\textwidth]{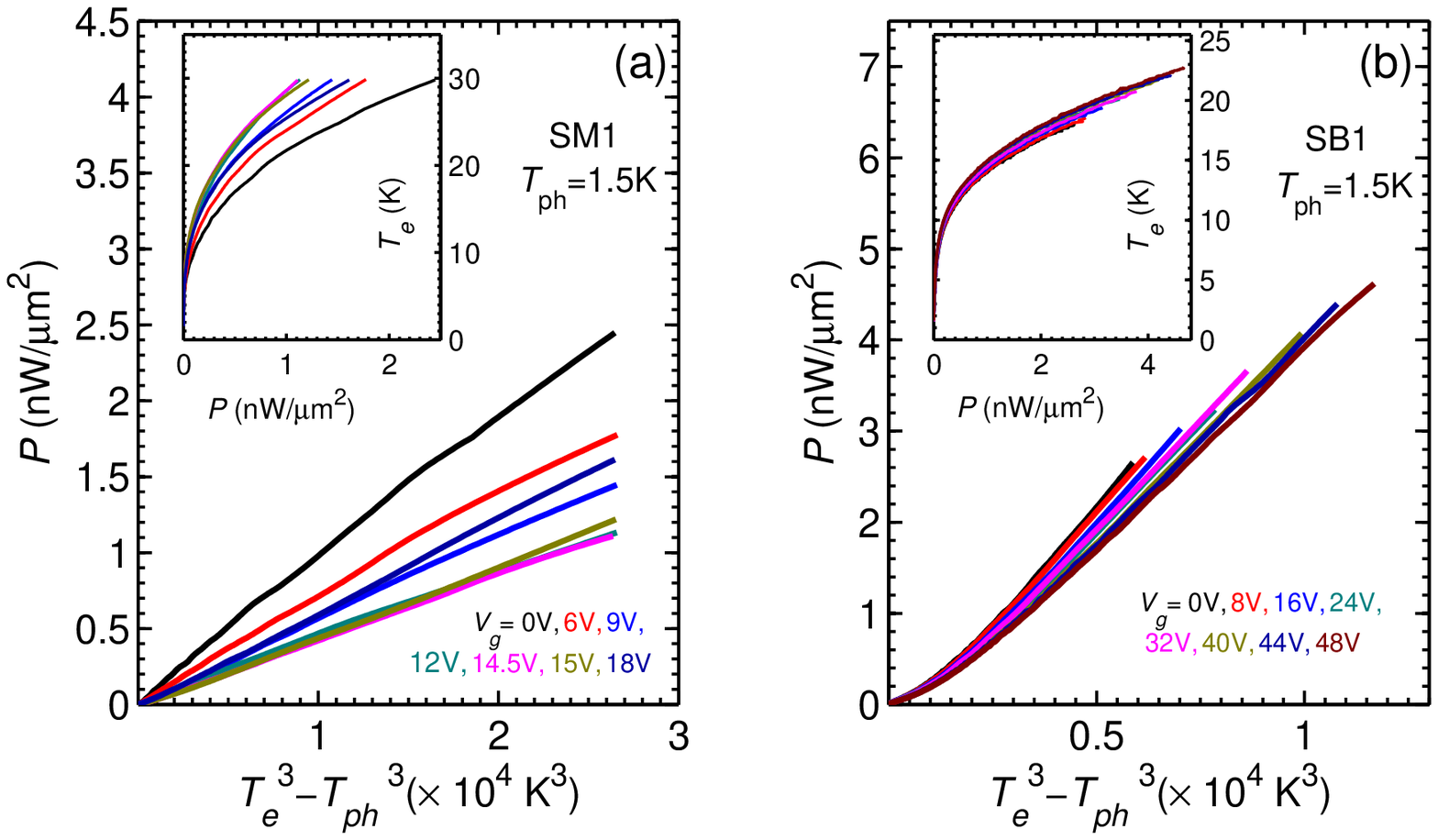}
\caption{\label{PT} Cooling power of monolayer and bilayer defected graphene. (a)(b) Cooling power $P$ against $T_\text{e}^3-T_\text{ph}^3$ shows a linear dependence for both monolayer and bilayer samples. Inset: $P$ versus $T_\text{e}$.}
\end{figure}

Upon Joule heating, the electronic temperature is raised, leading to decrease of the resistance, depicted in \rfig{rt}b. Based on the resistance as a function of temperature, we obtain the $P-T_\text{e}$ relation at different carrier densities, plotted in the insets of \rfig{PT}. $P$ is also plotted against $T_\text{e}^3-T_\text{ph}^3$. The linear behavior agrees well with \req{eq1} with $\delta=3$ for both monolayer and bilayer graphene at all carrier densities. It has been theoretically shown that both clean monolayer and bilayer graphene can be described by \req{eq1} with $\delta=4$ at low temperature \cite{Viljas2010,Kubakaddi2009}. In presence of disorder, e-p interaction is enhanced and $\delta$ is reduced to 3\cite{Song2012,Chen2012}. $\delta$ obtained in our result is consistent with these theories, indicating the effect of defects. $T^3$ dependence has also been reported in some other experiments. In the following, we will compare our results in detail with previous theoretical and experimental results.

The e-p interaction is usually considered in two distinct regimes, high temperature and low temperature. In normal metals, Debye temperature $\theta_\text{D}$ demarcates two regimes. Below $\theta_\text{D}$, the phase space of available phonons increases with temperature, while it becomes constant above it(all modes are excited). In graphene, because of its low carrier density, the Bloch-Gr\"uneisen temperature $T_\text{BG}$ becomes the relevant characteristic temperature. It is defined as $2k_\text{B}T_\text{BG}=2hck_\text{F}$. Here $k_\text{B}$ is the Boltzmann constant, $h$ the Plank constant, $c$ the sound velocity of graphene and $k_\text{F}$ the Fermi wave vector. $T_\text{BG}$ stems from the momentum conservation in e-p scattering. Because of it, when $T_\text{ph}>T_\text{BG}$, only a portion of phonons can participate in the process\cite{Fuhrer2010}. Considering the band structure of graphene, we have $ T_\text{BG}=2(c/v_F)E_F/k_B$ in monolayer graphene and $ T_\text{BG}=2(c/v_F)\sqrt{\gamma_1E_F}/k_B$ in bilayer graphene\cite{Viljas2010}. Here $v_F\approx10^6$ m/s is the Fermi velocity, $c\approx2\times10^4$ m/s and $\gamma_1 \approx 0.4$ eV is the interlayer coupling coefficient. Taking into account a residual carrier density $n_0\approx 4\times 10^{11}$ cm$^2$ due to charge puddles\cite{Li2011,Zhang2009}, it can be readily estimated that even at the CNP, $T_\text{BG} > 34$ K. It is much higher than $T_\text{e}=1.5$ K in our experiment. Consequently, we are well in the low temperature regime.

\begin{figure}[htb] 
\includegraphics[width=1\textwidth]{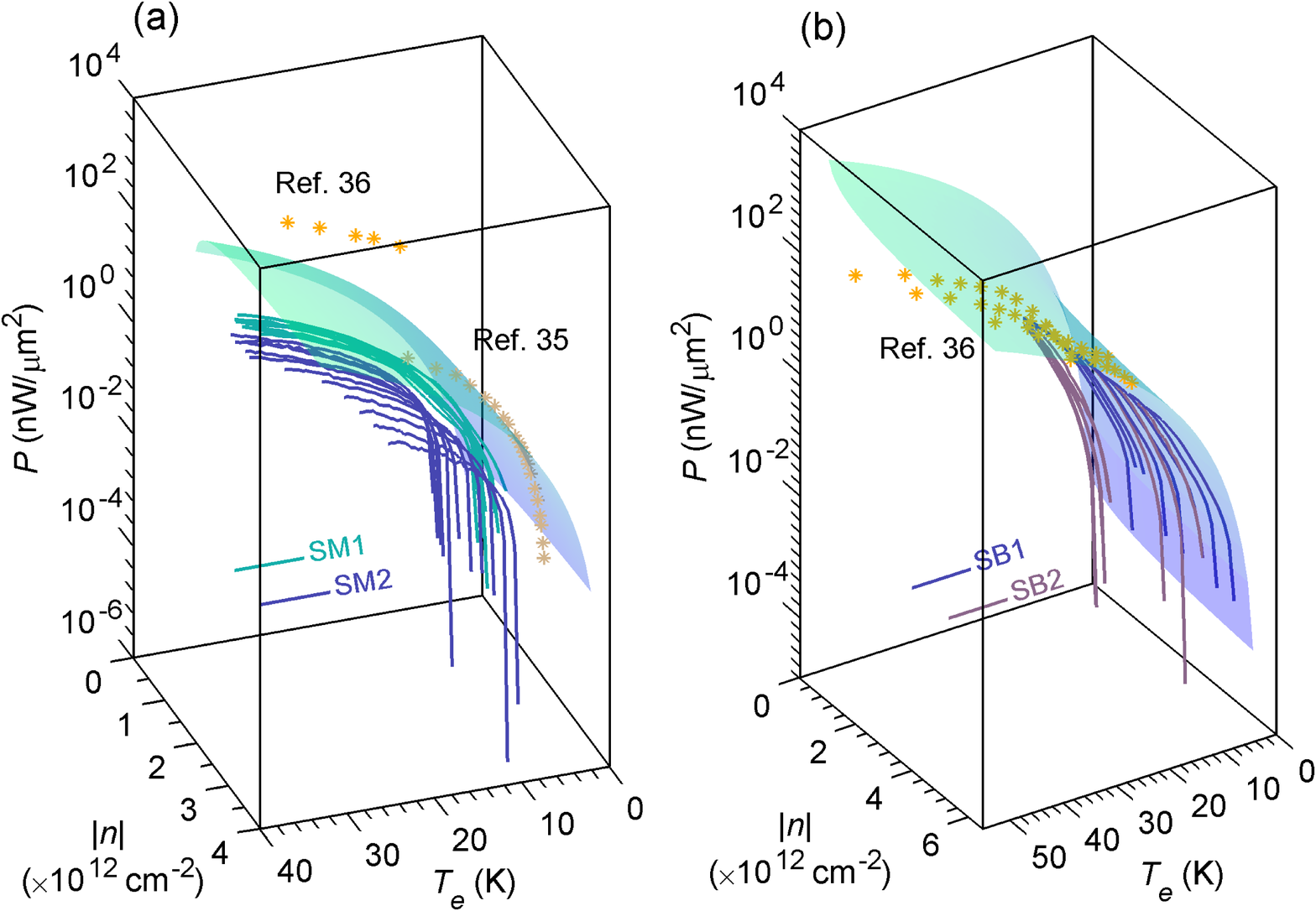}
\caption{\label{3d} Suppression of electronic cooling in defected graphene. (a) Cooling power of clean monolayer graphene is depicted as a transparent surface, with a logarithmic z-axis scale and as a function of $T_\text{e}$ and $n$. $P$ for SM1 and SM2 are over an order of magnitude smaller than clean graphene at all carrier densities, while the data from others' work is either on or above the surface. $n_0$ is chosen as 4 $\times 10^{11}$ cm$^2$ to account for charge puddles near CNP. $T_\text{ph}$ is 1.5 K in SM1 and the theoretical surface, 7K in SM2, 0.8 K in the data from \rref{Fong2013} and 1.8 K in the data from \rref{Somphonsane2013}. (b) Similar suppression is observed in bilayer defected graphene samples. $n_0$ is chosen as 4 $\times 10^{11}$ cm$^2$. $T_\text{ph}$ is 1.5K in SB1, SB2 and the theoretical surface, and 1.8 K in the data from \rref{Somphonsane2013}.}
\end{figure}

In the low temperature regime, the whole population of phonons can interact with electrons. Thus, the disorder-assisted supercollision is negligible\cite{Song2012}, which rules out it as the origin of the observed $T^3$ dependence. It has been theoretically shown that in the case of weak screening, static charge impurities leads to enhanced e-p cooling power over clean graphene and $\delta=3$\cite{Chen2012}. For comparison, we plot our data, the theoretical cooling power of clean graphene in \rfig{3d}. The theoretical prediction of the cooling power per unit area in clean monolayer graphene is \cite{Viljas2010}
\be
P_{\text{clean}}=\frac{\pi^2D^2 E_\text{F} k_B^4}{15\rho \hbar^5 v_\text{F}^3 c^3}(T_\text{e}^4-T_\text{ph}^4)
\ee
where $\rho \approx 0.76\times10^{-6}$ kg/m$^2$ is the mass density of graphene and $D$ is the deformation potential chosen as a common value 18 eV \cite{Chen2008a,Graham2013,Fong2013} (this choice will be discussed later). The theoretical cooling power $P_{\text{clean}}$ as a function of the carrier density and electron temperature is depicted as a transparent surface (with $T_\text{ph}$ =1.5 K) in \rfig{3d}a. It can be clearly seen that the cooling power of our disordered samples SM1 and SM2 (green and blue lines) are well below the surface at all carrier densities. For comparison, we also plot the data from two other experiments in which $T^3$-dependence were observed at low temperatures\cite{Fong2013,Somphonsane2013}. These results (with similar $T_{\text{ph}}$) are either on or above the surface. The suppression of the cooling in \rfig{3d}a is considerable. For instance, at $n=4\times10^{11}$ cm$^2$ and $T_{\text{e}}=20$ K, the theory predicts $P_{\text{clean}}$=4.7 nW/$\mu$m$^2$. In \rref{Somphonsane2013} the cooling power was found to be 27 nW/$\mu$m$^2$. In sharp contrast, our experiment gives a cooling power of 0.33 nW/$\mu$m$^2$ for SM1, over an order of magnitude lower than that in clean graphene. For the more disordered sample, SM2, it is even smaller.

Similar suppression occurs in bilayer graphene samples, too. The cooling power per unit area in clean bilayer graphene is given by \cite{Viljas2010}
\be
P_{\text{clean}}=\frac{\pi^2D^2 \gamma_1 k_B^4}{60\rho \hbar^5 v_\text{F}^3 c^3}\sqrt{\frac{\gamma_1}{E_\text{F}}}(T_{\text{e}}^4-T_{\text{ph}}^4)
\label{bilayer}
\ee
\rfig{3d}a shows the plot of \req{bilayer}, the cooling power of the bilayer samples SB1, SB2 and the data from \rref{Somphonsane2013}. Although not as pronounced as monolayer graphene, our data still below the theoretical surface. The weaker suppression may result from the fact that the bottom layer of bilayer graphene has experienced less damage by our low energy plasma than the top one\cite{Chen2013}. Therefore, this less disordered layer provides a channel of substantial cooling. 

The e-p coupling strength depends on the deformation potential $D$, which characterizes the band shift upon lattice deformation\cite{Bardeen1950,Herring1956,Suzuura2002}. 
For the theoretical cooling power surface in \rfig{3d}, we use $D=18$ eV. Note that $D$ for graphene ranges from 10 to 70 eV in various experiments, but 18 eV is the most common value for graphene\cite{Fong2013}. If the suppression is due to an over-estimated $D$, to account for the small cooling power, one would require $D$ to be only about 5 eV, one-half of the lowest value reported. Therefore, we believe that the suppression cannot be explained by a small $D$.

\begin{figure}[htb] 
\includegraphics[width=1\textwidth]{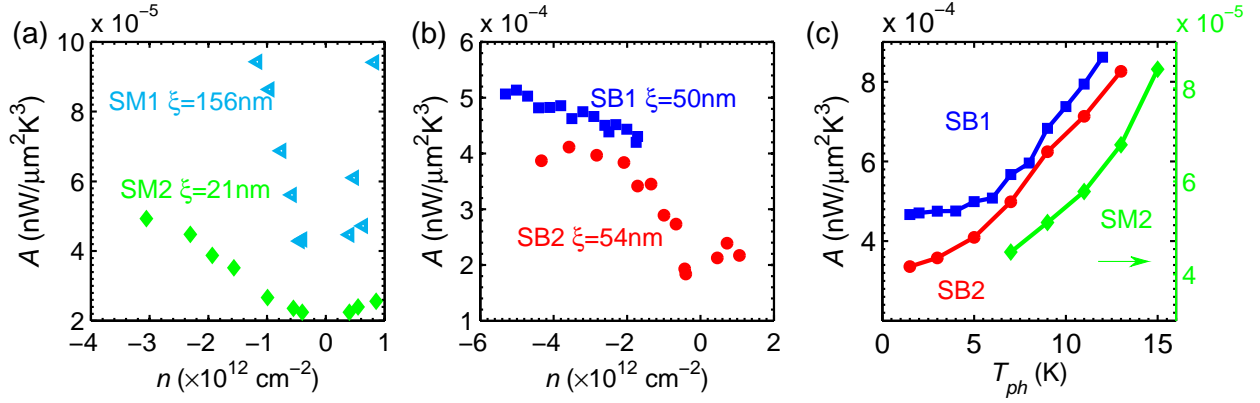}
\caption{\label{A} Coupling constant $A$. (a) Extracted coupling constant $A$ as a function of carrier density $n$ in monolayer samples. The more defective sample, SM1, exhibits a smaller coupling constant. (b) Dependence of $A$ on carrier density $n$ in bilayer graphene samples. The curves for two samples with similar degree of disorder align reasonably well. (c) Dependence of $A$ on phonon temperature $T_\text{ph}$. The data of SM2 are plotted with respect to the right y-axis.}
\end{figure}

By linear fits of $P$ versus $T_\text{e}^3-T_\text{ph}^3$, the coupling constant $A$ can be obtained. In \rfig{A}, $A$ is plotted as a function of carrier density $n$. $A$ for all samples decreases when approaching the CNP. This is because fewer carriers at Fermi level could contribute to total cooling power of the sample.

We now take a look at the dependence of the coupling constant on the degree of disorder. As listed in \rtab{tab:info}, the samples have been subject to various periods of plasma treatment. Consequently, the degree of disorder is different, indicated by the localization length $\xi$. For instance, $\xi$ for SM1 and SM2 is 156 nm and 21 nm, respectively. As plotted in \rfig{A}a, the coupling constant $A$ of the less disordered SM1 is only about one-third of the value for the more disordered SM2. The dependence of $A$ on $\xi$ is consistent with the suppression of the e-p scattering by disorder. For the two bilayer samples, SB1 and SB2, the localization lengths are close. The $n$ dependence of $A$ for both samples aligns reasonably well and is consistent with the monolayer samples, see \rfig{A}b.

The Joule heating experiment has also been carried out at different phonon temperatures $T_\text{ph}$. In \rfig{A}c, the coupling constant $A$ is plotted as a function of $T_\text{ph}$. Usually, $A$ is independent of $T_\text{ph}$, which is actually seen at low temperature for SB1. However, as the temperature goes above 7 K, $A$ is enhanced. Later, we will show that the unexpected $T$-dependence is likely related to the dynamic nature of vacancies.

At first glance, the suppression of electronic cooling by vacancies seems surprising, in that previous theories have predicted that disorder would enhance the cooling\cite{Chen2012,Song2012}. Most of earlier experimental results have confirmed the enhancement\cite{Somphonsane2013,Fong2013,Betz2013,Graham2013}. However, there is a key difference between those earlier studies and ours. In the former, disorder is theoretically considered to be static. This is indeed true in other experimental work, in which the dominant disorder is due to charge impurities\cite{Adam2007,Chen2008}. However, in our samples, the dominant disorder is vacancies, which are completely dragged by phonons. The effect of disorder on the e-p interaction has been studied in disordered metals and found to depend on the character of disorder\cite{Schmid1973,Sergeev2000,Lin2002,Zhong2002}. In the case of static disorder, diffusive motion of electrons increases the effective interacting time between an electron and a phonon, leading to an enhancement of interaction. However, dynamic disorder modifies the quantum interference of scattering processes\cite{Sergeev2000}. As a result, the interaction is suppressed, in accordance with the famous Pippard's inefficient condition\cite{Pippard1955}. It is reasonable to believe that the observed suppression results from dynamic disorder, vacancies. Furthermore, since the dynamics of disorder apparently depends on $T_\text{ph}$, the dependence of the coupling constant $A$ on the phonon temperature $T_\text{ph}$ is then conceivable. As described in Schmid's theory\cite{Schmid1973,Sergeev2000}, the e-p scattering is suppressed due to strong disorder. The resultant energy relaxation rate $\tau_\text{e-p}^{-1}$ is of the order of $(q_T l) \tau_0^{-1}$ where $\tau_0^{-1}\propto T^3$ is the relaxation rate in pure material, $q_T$ is the wave vector of a thermal phonon and $l$ is the mean free path. As $q_T \propto T_\text{ph}$, the relaxation rate increases with $T_\text{ph}$, in agreement with our result.

It is also worthy to note that charge impurities are long range potentials that preserve the sublattice symmetry. This is in contrast to vacancies, which are short range potentials and break the sublattice symmetry. The theory for supercollision models disorder as short range potential\cite{Song2012}, while in \rref{Chen2012}, disorder potential is long-ranged. This character of disorder strongly affects scattering of chiral electrons in graphene. Our samples represent a graphene system that is quite different from what was commonly seen, in that dynamic and short-ranged potentials dominate. Therefore, the quantitative understanding of our experimental results, including the power index $\delta$, relies on future theory that takes both the dynamics and the symmetry of disorder into account. 

\section{Conclusion} 
In conclusion, we have observed significant suppression of electronic cooling in defected graphene. The cooling power of both monolayer and bilayer graphene samples show $T_\text{e}^3$ dependence, consistent with disorder-modified electron-phonon coupling in graphene \cite{Chen2012,Song2012}. However, the magnitude of the cooling power is over an order of magnitude smaller than that of clean graphene predicted by theory\cite{Kubakaddi2009,Viljas2010} and also less than other experiments \cite{Fong2013,Somphonsane2013}. The more disordered a graphene film is, the lower cooling power is observed, confirming the effect of disorder. The suppression of electronic cooling is attributed to the dynamic nature of vacancies, which has not been studied in graphene. This effect can be utilized to further improve the performance of graphene-based bolometer and photo-detector devices. 

\begin{acknowledgments}
This work was supported by National Key Basic Research Program of China (No. 2012CB933404, 2013CBA01603) and NSFC (project No. 11074007, 11222436, 11234001).
\end{acknowledgments}


\newpage

\clearpage

\end{document}